\documentclass{article}

\usepackage{PRIMEarxiv}

\usepackage[utf8]{inputenc} % allow utf-8 input
\usepackage[T1]{fontenc}    % use 8-bit T1 fonts
\usepackage{hyperref}       % hyperlinks
\usepackage{url}            % simple URL typesetting
\usepackage{booktabs}       % professional-quality tables
\usepackage{amsfonts}       % blackboard math symbols
\usepackage{nicefrac}       % compact symbols for 1/2, etc.
\usepackage{microtype}      % microtypography
\usepackage{lipsum}
\usepackage{fancyhdr}       % header
\usepackage{graphicx}       % graphics
\graphicspath{{media/}}     % organize your images and other figures under media/ folder

\title{More Than a Wife and a Mom: A Study of Mom Vlogging Practices in China}

\author{
  Kyrie Zhixuan Zhou\textsuperscript{1},
  Bohui Shen\textsuperscript{2},
  Franziska Zimmer\textsuperscript{3},
  Chuanli Xia\textsuperscript{2},
  Xin Tong\textsuperscript{4} \\
  \textsuperscript{1}University of Illinois at Urbana-Champaign\\
  zz78@illinois.edu\\
  \textsuperscript{2}BNU-HKBU United International College\\
  r130233082@mail.uic.edu.cn, chuanlixia@uic.edu.cn\\
  \textsuperscript{3}Heinrich Heine University Düsseldorf\\
  franziska.zimmer@hhu.de\\
  \textsuperscript{4}Duke Kunshan University\\
  xin.tong@dukekunshan.edu.cn\\
}

\begin{document}
\maketitle

\begin{abstract}
  Mom vloggers are stay-at-home moms who record and share their daily life through short videos. 
  % Though mom vlogging enables them to have an income and seek their identities and values, challenges abound such as degraded privacy of family members. 
  In this exploratory study, we aspire to understand mom vloggers' motivations, practices, and challenges. Our mixed-methods inspection contained interviews with 4 mom vloggers in China and a content analysis of mom vlogs of 5 other mom vloggers. Mom vloggers' primary motivations are to make money, record daily life, and seek their individual identities and values, well meeting their financial and social needs after leaving their paid employment. 
  % Revenues gained from advertising/selling products may make up for the decrease of income after leaving their paid employment, and together with the support and feedback from viewers, our participants' worth-based self-esteem is fulfilled. 
  When creating vlog content, mom vloggers encounter various challenges, such as a lack of video visibility, being stretched by both intensive motherhood and heavy digital work, privacy and self-presentation concerns, and so on. Based on the findings, we propose design implications toward resolving these challenges and benefiting mom vloggers' experiences.
\end{abstract}

% keywords can be removed

\section{Introduction} 

\begin{figure*}[h]
\centering
  \includegraphics[width=0.7\textwidth]{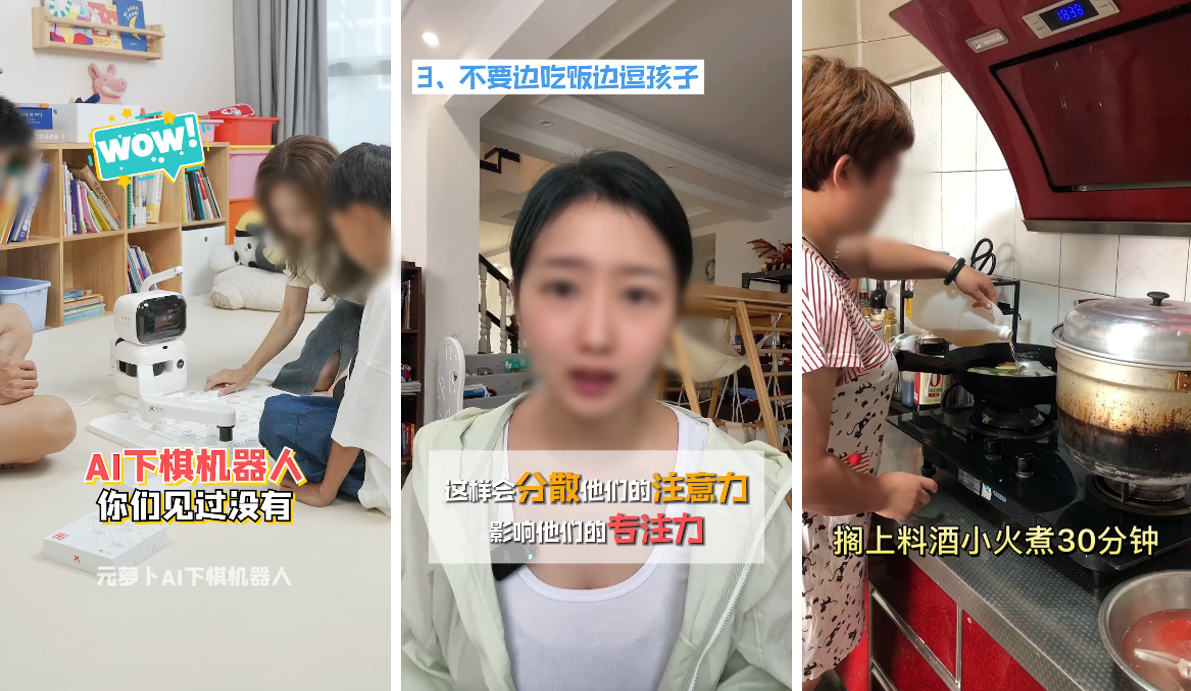}
  \caption{Example screenshots of typical mom vlogs. From left to right: promoting an AI chess-playing robot (C1), sharing parenting tips in feeding kids (C3), and recording daily cooking (C4).}
  \label{fig:teaser}
\end{figure*}

Recently, mom vlogs are emerging on video-sharing platforms in China, such as Douyin\footnote{\url{https://www.douyin.com/discover}} and Xiaohongshu\footnote{\url{https://www.xiaohongshu.com/}} \cite{mom:vlogging}. In the vlogs, the new moms record their daily life after giving birth (e.g., 
housework and childrearing), provide parenting and education tips to other moms, advertise products, and so on (see Figure \ref{fig:teaser}). 
Despite the popularity of mom vlogging, the vloggers' motivations and challenges have not been sufficiently investigated yet. 

New parents, especially moms, often suffer from loneliness due to a reduced social life arising from heavy family duties and financial difficulties \cite{lonely}. According to a survey of 529 parents of young children in the UK, 56\% of the parents feel lonely and see children-themed group activities as a potential way to resolve their loneliness \cite{lonely:2}. Stay-at-home moms can be a more vulnerable population. 
Even highly-educated women in the UK have to leave paid employment to take care of their children, demonstrating the gap between the promise of gender equality and women's experience of continued injustice \cite{heading}. 

The situation for moms is even worse in China, where our study has been conducted, with more traditional gender roles assigned to women~\cite{widow,anonymous}.  
In China,  21.8\% of Generation Z moms do not have paid jobs ~\cite{stay}. Stay-at-home moms often face difficulty in terms of re-employment due to the exhausting childrearing burdens and a lack of legal protection \cite{housewife:1,housewife:5,housewife:3}. The considerable income of popular streamers \cite{popular} and social opportunities afforded by short videos (e.g., comments, fan groups, etc.) \cite{streaming:3} have the potential to accommodate some of the stay-at-home moms' financial and social needs \cite{lonely}, though challenges such as degraded privacy for family members \cite{mom:6} and cyberbullying targeting streamers \cite{cyberbullying} arise.
 
Based on these considerations, our research questions (RQs) are:
\begin{itemize}
    \item RQ1: Why do moms decide to become vloggers and post their life experiences on video-sharing platforms?
    \item RQ2: What types of content do they generate?
    \item RQ3: What challenges do mom vloggers encounter and what are their coping strategies?
\end{itemize}

To understand mom vloggers' motivations, practices, and challenges, we conducted an exploratory study combining semi-structured interviews and qualitative content analysis. 
Our interview results suggested that making money, recording life, and seeking their identities and values were the main factors motivating moms to become vloggers. 
Though stay-at-home moms experienced an enriched life through vlogging, they encountered challenges in: (1) gaining more visibility on social platforms, especially for novice vloggers, (2) managing efforts and time regarding motherhood duties and digital work, (3) facing personal attacks due to competition among moms, and (4) other miscellaneous aspects, e.g., protecting privacy and managing self-presentation. Content analysis results of the mom vlogs validated the prior interview findings and also identified the diverse and evolving topics of mom vlogs. 

Our empirical research strengthened HCI communities' understanding of Chinese mom vloggers' practices and challenges with design implications, which could potentially benefit mom vloggers' experiences and inspire future research in ICT for moms.

\section{Related Work}

\subsection{Short Videos}
Short videos, like those shared on TikTok, are rapidly gaining attention in recent years \cite{streaming:5}. TikTok was the most downloaded app in 2020, with 850 million downloads globally \cite{streaming:6}. The huge live-streaming industry in China has been influenced by short video sharing mobile applications \cite{streaming:7} and researchers have investigated various short video genres, e.g., tourism \cite{tourism} and food \cite{streaming:1}. In this work, we focus on a specific short video genre, i.e., mom vlogging, which is not well understood so far.   

Several issues and challenges are associated with short videos and streaming. Exploitation of labor of Kuaishou streamers was reported in \cite{mom:1}: the app takes half of all the gifted money that a streamer receives, before transmitting the remainder to them. Thus these streaming apps simultaneously empower streamers with opportunities for money and social recognition, and disempower them with exploitation \cite{mom:2}. There has always been a compromise between self-disclosure and self-presentation when it comes to social media usage. While the anonymity of the Internet encourages people to disclose their actual selves, other features of it, such as asynchronicity, multiple audiences, and published audience feedback, have discouraged disclosure \cite{present:1}.

% Exploitation of labor of Kuaishou streamers was reported in \cite{mom:1}: the app takes half of the gifted money from streamers. The platforms thus simultaneously empower streamers with opportunities of money and social recognition, and dis-empower them with financial exploitation \cite{mom:2}. 
% There has always been a compromise between self-disclosure and self-presentation when it comes to social media usage. While anonymity of the Internet encourages people to disclose their actual selves, other features of it, such as asynchronicity, multiple audiences, and published audience feedback, have discouraged disclosure \cite{present:1}. 
% On the one hand, people expect that disclosing more online leads to positive outcomes, such as the growth of close relationships \cite{present:2}. On the other hand, social media users often tend to present an edited version of themselves that they believe will be best received by others \cite{present:3}.

\subsection{Mom Blogging}
While our investigation focuses on mom vlogging, a relevant media is mom blogging in non-video forms \cite{mom:3}. 
Researchers have coded mom influencer accounts on Instagram and identified diverse emotions and topics in the posts \cite{meso}. 
Australian blogging mothers were found to negotiate their identities as mothers and move beyond their homes using social media, in which process they made money and obtained a sense of belonging \cite{mom:3}. A survey study revealed mothers' motivations for blogging were to develop connections with others, achieve self-validation, contribute to the welfare of others, and extend skills and abilities \cite{mom:4}. Privacy concerns, however, are inevitable. For example, ``Mumpreneur bloggers'', who earned a significant income from their personal brand and role as social media influencers, may use their family members, particularly their young children, as characters in their story posts, potentially leading to privacy leakage \cite{mom:6}. The visual nature of mom vlogging can make vloggers more vulnerable to privacy risks. For example, some YouTube kids' lives have been documented in vlogs from the moment they were born \cite{youtube}. 

\subsection{Gender, Motherhood, and Feminist HCI}

Intensive motherhood has long been a gender issue in China \cite{housewife:1,housewife:2,housewife:3,housewife:4,housewife:5}. Women need to seek a delicate balance between childrearing and their own work and life. They also have to face the conflict between their own perception of the ideal status of being a mother and motherhood expected by the family and society. The gender division within the family and the dominant motherhood ideology outside the family make the conflict harder \cite{gender:2}. 
% Hanser and Li argued that intensive, demanding forms of parenting extended into the earliest years of a child's life, and suggested that the linkage between breastfeeding and motherhood represented a ``gendered burden'' for Chinese women \cite{gender:1}.
Many mothers found themselves increasingly isolated and overwhelmed after giving birth to a
new baby, leading to extreme stress, anxiety, and depression \cite{mom:5}.

ICT has a long history of being used to help solve gender issues. While women have been historically disadvantaged in careers compared to men, technology brought new jobs for them, e.g., web-girl. It not only opened door for self-presentation but also enabled women to seek different roles as role makers \cite{web}. 
The success of Xiaohongshu, a Chinese social media platform, has been known to stem from its emphasis on female consumption \cite{xiaohongshu}. Its cultural concept, media convergence, and marketing are highly connected with feminism.

The value of mothers' labor \cite{labor} and labor support for mothers \cite{labor:2} have been extensively studied. However, gender-related social protection in China is overshadowed by a conservative gender culture, with some employers refusing to be compliant with paid maternal leave and paternity leave not being included in the Chinese legal system at all \cite{labor:3}.
Through the lens of mom vlogging, we aspire to uncover the challenges faced by Chinese mothers.

\section{Method}
We first conducted semi-structured interviews with mom vloggers to have an overall understanding of their motivations (RQ1) and challenges (RQ3), followed by a qualitative video content analysis of a different set of mom vloggers to further understand the content of their vlogs (RQ2) and validate the interview results.

\subsection{Exploratory Interviews}
\noindent \textbf{Participants.}
We sent direct messages to mom vloggers on Douyin and Xiaohongshu to recruit participants, after searching the keyword ``mom vlogger'' (\begin{CJK}{UTF8}{gbsn}宝妈主播\end{CJK}). We watched mom vlogs to identify potential interviewees as well as to familiarize ourselves with this video genre.
In the end, four mom vloggers responded to us after we sent 100 recruitment messages.

The demographic information of these participants was summarized in Table \ref{tab:table1}. All but P1 had more than 10k followers, and their experience of mom vlogging ranged from 1 month to more than 1 year. P2 was a teacher on maternal leave, and all other participants were stay-at-home moms at the time of this study.  
 
\begin{table*}[htb]
  \centering\resizebox{0.8\textwidth}{!}
  {\begin{tabular}{l r r r r r r}
    % \toprule
    {\small\textit{ID}}
    & {\small \textit{Age}}
    & {\small \textit{Platform}}
    & {\small \textit{Education}}
    & {\small \textit{Number of followers}}
    & {\small \textit{Prior occupation}}
    & {\small \textit{Mom vlogging experience in month(s)}}
    \\
    \midrule
    P1 & 30 & Douyin & Bachelor & 148 & Taobao streamer & 1\\
    P2 & 28 & Douyin & Master & 35k & Teacher & 9\\
    P3 & 33 & Xiaohongshu & Master & 12k & Sport shop manager & 15\\
    P4 & 33 & Douyin & Associate & 33k & Business & 5\\
    \midrule
    C1 & - & Douyin & Master & 10M & - & 57\\
    C2 & 28 & Douyin & Bachelor & 1M & - & 16\\
    C3 & 28 & Douyin & - & 110k & - & 14\\
    C4 & 37 & Douyin & - & 35k & - & 24\\
    C5 & 38 & Douyin & - & 1k & - & 48\\
    % P5 & ? & Xigua & ? & 131 & ? & ?\\
    % \bottomrule
  \end{tabular}}
  \caption{Demographic information of mom vloggers as of Apr. 2022 for interviewees (P1-P4) and Dec. 2022 for content creators in the content analysis (C1-C5).  Age, education, and prior occupation information is not available for some content creators.}~\label{tab:table1}
\end{table*}

\noindent \textbf{Semi-structured Interviews.}
We started the interviews by introducing our study intentions, requesting participants' consent, and collecting their demographics. We then went on to ask open-ended, in-depth questions about their vlogging experiences, 
such as attitudes toward privacy leakage, feedback from viewers, income changes, and so on. The full interview protocol can be seen in Section \ref{interview}.

One author who is a native Chinese speaker conducted the interviews via audio calls between April and May 2022. 
Each interview lasted around 30 minutes, and was audio-recorded and auto-transcribed with a tool provided by Sogou\footnote{\url{https://rec.sogou.com/voice}}.
The research team then checked the transcripts and corrected errors.

\noindent \textbf{Data Analysis.}
We adopted a thematic coding approach for analysis \cite{thematic}. Two authors independently coded the data and regularly discussed to reach a consensus. We used mind mapping to organize the emerging themes, subthemes, and quotes into a hierarchical structure. We will use anonymized quotes to illustrate our findings.

\subsection{Vlog Content Analysis}
\noindent \textbf{Data Collection.}
We collected 50 mom vlogs created by five vloggers in Douyin. We specifically sought vloggers with different levels of fan bases to capture a comprehensive picture of the vlogs' content. These mom vloggers had 1,000 to 10M followers (see C1-C5 in Table \ref{tab:table1}). 
We collected their very first five vlogs and the most recent five vlogs to understand how their vlogs have evolved in terms of topics, quality, and interaction dynamics in the comments. 

\noindent \textbf{Data Analysis.}
We focused on analyzing the videos' topics, quality, motivations, and challenges if revealed. Two authors adopted the four-eyes principle \cite{principle} to observe videos simultaneously but independently, and communicate afterward. 
Disagreement may occur, but consensus was always reached after discussion. We also looked into the videos' comments informally. 

\section{Findings from Interviews}
Through the analysis, we identified two main themes in the interviews, namely, heading home and mom vlogging. These two themes were highly related in nature: heading home was the reason for mom vlogging for all our participants; had they not headed home, they would not have started their mom vlogging careers. Thus, we will first report why they headed home and what stay-at-home mom life was like to contextualize later results on mom vlogging practices.

\subsection{Heading Home}
\label{heading:home}
Shani Orgad's book, \textit{Heading Home: Motherhood, Work, and the Failed Promise of Equality}, told the stories of how highly educated London women had to leave paid employment to take care of their children, while their husbands continued to work in high-powered jobs \cite{heading}. We hereby borrow the ``heading home'' notion and examine why our participants choose to head home as well as what their life is like after heading home.

\subsubsection{Why Heading Home?}
Our participants chose to head home and become stay-at-home moms for several reasons, from accompanying their children (N=4) to self-identifying as better caregivers than their partners (N=4). 

\textbf{Accompanying Children.}
P3 did not want to miss her child's growth, especially the first few years of their life. 
She mentioned that the maternity leave was too short (128 days), and she might not quit her job if it was longer. 
Notably, P1 told the authors that her husband only had 15 days of paternity leave.
The difference in the duration of maternity and paternity leave \cite{maternity:leave} revealed the social expectation of mothers as main caregivers.

\textbf{``Women Are Better Caregivers.''}
Interestingly, women were stereotypically perceived as ``better caregivers'' than men by our participants. For example, P3 acknowledged that she was not as ambitious as her husband in career development and treated heading home as a natural and reasonable decision. She elaborated, \textit{``In my family, it's obvious that I like to take care of children more than my husband. He wants to achieve high in his career, but I'm never such an ambitious person. So, it's a natural choice for me [to head home].''}

\subsubsection{What it's Like Being A Mom Vlogger?}
Quitting jobs and becoming stay-at-home moms have inevitably brought changes and challenges to our participants' lives (N=4), among which reduced income was repeatedly mentioned (N=3). Other aspects included intensive labor for housework and reduced social life.

\textbf{Reduced Income.}
The revenue of novice vloggers was hardly comparable to their previous paid employment before they could successfully make profits from advertising and selling products. The reduced income has made P1 feel insecure about her family's ability to face risks. Such financial challenges motivated her to gain more visibility and attract more followers, and ultimately, earn more money.

\textbf{Intensive Labor.}
All participants noted that housework was labor-intensive throughout the day, and none of them mentioned their husbands' roles in housework and parenting. When we interviewed P1, she was interrupted by the crying of her baby several times. She narrated on her daily routine, \textit{``I get up at 6 in the morning, make breakfast for my family, and send my kid to school. Then I do the housework, design, shoot and edit videos, and pick up my kid from school. After playing with my kid outside, I make dinner, edit and post videos, and interact with my followers.''}

\textbf{Reduced Social Life.}
Our participants experienced reduced social activities after giving birth. P3 shared, \textit{``My friends are mostly coworkers from my previous job. I can't see them now, so I have less social life.''} The reduced income and social life after heading home have motivated some moms to start their vlogging careers, which we will elaborate on shortly. 

\subsection{Fulfilling Yet Challenging Mom Vlogging}
Our participants' vlogs were mostly about housework (P1), pregnancy (P2), education (P3), family (P3, P4), and product recommendations (P2). All of them had an overall good experience with mom vlogging and viewed it as a job or career (N=4). They enjoyed a sense of identity and achievement through sharing skills (P2), encouraging novice moms (P1, P2, P3, P4), and providing parenting and education suggestions (P2, P3). 

\subsubsection{Motivations for Becoming a Mom Vlogger.}
Recording life as new moms, from cooking to childrearing, was a common motivation for our participants to start their mom vlogging careers (N=4). 
Other motivations included relevant work experience (N=2), making money (N=3), and seeking identities and values (N=4).

\textbf{Relevant Prior Working Experience.}
Some participants' and/or their partners' relevant working experiences have enabled them to initiate and succeed in their vlogging careers. For example, P1 had been an E-commerce streamer on Taobao before she had a child. In another case, P2 was a teacher on maternity leave, who had strong presentation and communication skills. Her husband was a staff at Douyin who encouraged her to become a mom vlogger and provided her with inner insights. 

\textbf{Making Money.}
The vloggers who had a relatively large fan base would advertise products on streaming platforms. 
Vloggers shared 10\% to 35\% of the profits (P1, P2) with the vendors who reached out to them. The platform would chip in and take 6\% (P1) or 2\% (P2) of their income, which was perceived as acceptable by them. P4 would further use live streaming to boost sales, since direct interactions with the audience would encourage them to buy more, according to her. P2 used fan chat groups on WeChat as an additional channel to sell products, provide customer services, and build a community. 

\textbf{Community and Self-value.}
Notably, all participants actively sought an online community to identify with and realize their values. 
For instance, P4 said, \textit{``I want to realize my true value other than being a good wife and mom.''} 
Similarly, P3 told us that moms would socialize with other moms to build a community and encourage each other, \textit{``By being a vlogger, I'm not trapped in the family anymore. Before, it seemed the only things in my life were my husband and my kid... Moms would look for such a group and encourage each other.''}

\subsubsection{Challenges and Mom Vloggers' Coping Strategies.}
Our participants (N=4) perceived no technical challenges when creating videos.
The challenges reported by them were the visibility of their vlogs, conflicts between intensive motherhood and digital work, privacy and self-presentation, personal attacks, and platform regulation. 

\textbf{Striving for More Views.}
Gaining more views was a common challenge among our participants (N=4), especially for novice content creators like P1, who had relatively few followers. The number of views received by each video was unpredictable due to the black-box nature of recommendation systems \cite{black}. For example, P2 complained that, often, videos that she perceived as of high quality were not seen by many viewers, which made her frustrated, \textit{``As a teacher, my outcome is controllable. If I'm more responsible, I can directly see the improvement in my students' grades. But Douyin is so unpredictable, and you can hardly succeed on your own. You think your videos are good and heart-touching, but there are just no likes or comments. So, it's frustrating sometimes.''} 

Our participants showed different strategies for gaining more views. In addition to improving content quality, the relevance of content was seen as a key to attracting views (N=4). For instance, P1 tended to post about issues that people may often come across in their daily lives to facilitate more interaction with the audience. 
For P4, the secret was presenting unique content that other mom vlogs did not afford, \textit{``In addition to good luck and perseverance, another important thing is how your content is different from other vloggers'. You can get up early and make a lot of delicate dishes for breakfast. I can do that as well, but I don't want to waste food. So, my breakfast videos receive fewer views.''} Controversial topics or negative experiences may also help expose videos to a broader audience but were not preferred strategies by our participants. For example, P4 explained, \textit{``I just want to share the happiness and `positive energy' in life. I don't want to give myself up for some [financial] purpose.''}

\textbf{Intensive Motherhood and Digital Work.}
Keeping a balance between intensive motherhood and heavy digital work is a major challenge for our participants (N=4). As mentioned in Section~\ref{heading:home}, our participants had to do housework and child care throughout the day. Producing vlogs was also labor-intensive: it took P4 two to three hours to make each vlog. P2 described how mom vlogging and interacting with followers/viewers overwhelmed her: \textit{``I treat it as a job, and I feel obligated to be responsible. So, I gotta check viewers' comments frequently. It took up much of my time.''}

The participants' coping strategies were consistent, i.e., prioritizing parenting and only doing mom vlogging in their free time (N=4). P3 sometimes used text posts to substitute video clips, which were more time-consuming. She explained, \textit{``I sometimes don't want to post videos because they take up too much of my time. I'd prefer posting in text... I have a schedule for each day, and I won't leave my child alone for a consecutive three hours.''} Outsourcing part of the vlogging or housework was a strategy shared by P2 and P3, who were in relatively good socioeconomic status. For instance, P2 hired a part-time assistant to edit the videos she recorded. 

\textbf{Privacy and Self-presentation.}
Our participants displayed varied attitudes toward privacy and self-presentation. P1 did not want her acquaintances to find her vlogging account, at least ``before her career was a success.'' 
She posted about her family life selectively, and would not share negative experiences, such as quarrels with her husband. She thought it ``immoral to solve family issues by posing the pressure of public opinion on family members.''

Other participants showed fewer concerns about privacy violations or self-presentation.
For example, P2 saw mom vlogging as a way of creating a ``bigger and broader friend circle,'' and for her, the financial gain outweighed privacy concerns. P3 thought her online image was not different from what she was like offline, so she did not worry about being seen by acquaintances. 

\textbf{``Mommy Wars.''}
Personal attacks were sometimes targeted at mom vloggers or their children and family (N=3), which negatively affected their feelings. 
Viewers may question the way mom vloggers raised their children, as observed by P3, \textit{``
In one video, my son ate noodles by himself.
Many moms criticized me for letting such a little kid feed himself. `He would choke.' `The parent is irresponsible.' Something like that.''} She saw this as a competition of moms, or ``mommy wars'' \cite{war,war:2}, where they tried to be the best in parenting. 

Some participants took a blunting approach \cite{monitor} and ignored the attacks. For instance, P2 said, \textit{``We're adults. I don't think it necessary to be affected by strangers' remarks.''}
P1, on the other hand, directly replied to the Douyin users who criticized her video quality,\textit{ ``I told them that there's not a set rule for everyone to behave well.''}

\textbf{Over- and Under- Regulation.}
Several issues were raised by P3 regarding platform regulation. She noted there were many clickbait vlogs teaching mothers how to take care of their babies ``easily.'' She thought such content could lead to peer pressure and anxiety for novice moms, and should be better regulated by the platforms. 

Censorship was also an annoying issue for her, \textit{``One time, I posted a breast-feeding picture on Xiaohongshu. I didn't show any part of my body, but I was warned of violating the community rule.''} She thought the sanction of motherhood content was ``ridiculous.''

\section{Findings from Vlog Content Analysis}

\subsection{Diverse and Evolving Vlog Quality and Topics}
We divided vlog quality into three levels, i.e., low, average, and high. We rated the vlogs based on how well-designed, well-shot, and well-edited they were. A high-quality video, for example, would tell a coherent story, capture the vlogger and what they do clearly, and leave redundant content out. The video quality of the 3 more popular vloggers was rated high, while the 2 less popular vloggers produced videos of average quality. Improvement was seen over time: C1's early videos were of average quality, but her recent ones were well-designed and of high quality. C5's early videos were of low quality with no substantial content; but later, her videos were improved to be of average quality. Video duration also increased over time (C1, C3, C5), from several seconds to several minutes, indicating vloggers' progressively advanced skills in content creation and video editing. 

A diverse range of topics were discovered in the videos, from parenting to education to cooking (e.g., infant food). Some mom vloggers changed their major video topics as their skills progressed or as they became more popular. For example, C1 initially created videos featuring cooking; later, as an influencer with millions of followers, she mostly provided parenting tips to help new moms. Similarly, C3 started out sharing pregnancy and childbirth experiences, and later her videos also featured parenting tips as her child grew up. Such changes revealed the trajectory of motherhood.

\subsection{Comments: A Mixture of Support and Attacks}
Followers' support was often seen in the comments. 
Some showed interest in the advertised products, possibly to boost the sales for the mom vloggers. Some uttered praise of the child's appearance and language ability, or the mom's culinary skills and parenting tips. Moreover, followers tried to offer useful suggestions for video creation, such as adding subtitles. Meanwhile, personal attacks such as the questioning of mom vloggers' parenting styles also existed in the comments, echoing the interview results. When C1 shared a video featuring making ice cream, many followers questioned that children would have diarrhea from eating cold food.

\section{Discussion}
With an exploratory study of Chinese mom vloggers, we uncovered their motivations, practices, and challenges of mom vlogging. Our participants become mom vloggers to record life, earn money, and seek their identities and values, echoing mom bloggers' motivations \cite{mom:4}. Some unique challenges abound compared to existing research on mom blogging in developed countries \cite{mom:3}. For example, our participants' partners hardly take childrearing responsibilities \cite{widow,anonymous}, leaving them overwhelmed by both intensive motherhood duties and heavy digital labor. Other challenges related to mom vlogging include vlog visibility, privacy concerns, etc. Next, we discuss motherhood in China and propose design implications to address mom vloggers' challenges, toward creating a better social channel for them to express and realize themselves. 
% especially in cultures where the value of housewives and stay-at-home moms is not well recognized \cite{labor:3}. 

\subsection{Motherhood in China}
The intensive labor required from stay-at-home moms makes them physically and mentally exhausted, and they can hardly spare time and effort to take care of themselves \cite{housewife:2}. Despite the time and energy they devote to taking care of and cultivating their children, when it comes to divorce and fighting for custody, they may not necessarily have an advantage due to their limited financial ability to raise children, even if they are well-educated women \cite{housewife:3}.

Several factors contribute to the challenges faced by stay-at-home moms. Childrearing responsibility, financial burden, and the conflict between work and family after giving birth have a profound impact on their re-employment process, regarding career choice, occupational status acquisition, and career mobility \cite{housewife:1,housewife:5}. A lack of support from husbands and discriminatory age requirements of jobs effectively marginalize Chinese mothers. Some regions have proposed the employment model of ``Mother's Post'' to promote women's employment, where the government financially supports employers to hire moms. However, experts have pointed out that such policies may solidify the traditional gender division of labor and reinforce the stereotype that mothers should take more responsibility in childrearing \cite{housewife:4}.

Our findings echo many of the challenges faced by Chinese moms, such as a lack of enforcement of paid maternity leave, limited support from husbands in terms of childrearing and housework, reduced social life, etc. Through mom vlogging, our participants were able to apply their professional skills (e.g., streaming) to their current careers, have an income, and identify a community where they have a sense of belonging.

\subsection{Mom Vlogging Challenges and Design Implications}
Improving video visibility is a major challenge, especially for novice mom vloggers. The black-box nature of recommendation systems makes it even harder for mom vloggers to speculate how much attention their vlogs would receive \cite{black}. 
Our participants shared strategies, such as featuring topics other moms could easily identify with and differentiating their topics from other vloggers. While popular mom vlogs often featured topics like parenting tips and infant/child product recommendations, 
platform designers could consider recommending trendy topics to novice mom vloggers when they create videos. 

Motherhood and digital work both require intensive labor work. 
While automatic video templates tailored for such scenarios as travel and anniversaries are already common \cite{douyin}, video-sharing platforms can further provide templates customized for mom vlogs.

Privacy concerns are inevitable in mom vlogging practices, especially when family members appear in the vlogs \cite{mom:6, attack, youtube}. Our participants held varied perceptions of privacy. For those who prefer to selectively post about their family life to protect their privacy or for self-presentation \cite{present:1}, future work could implement design features based on computer vision techniques to blur/hide family members who are unintentionally captured in the videos. 

Platform regulation can be improved to avoid over-censorship of motherhood content and under-regulation of inappropriate content and toxic comments. On one hand, posting breast-feeding pictures was seen as natural by moms but was not allowed under the strict social media censorship in China \cite{censorship:china, censorship:weibo}. On the other hand, clickbait emphasizing ``easy childrearing'' and bringing anxiety and peer pressure to new moms, as well as unwanted ``mommy wars'' \cite{war,war:2}, are not well regulated and could be misleading. 
In the future, platforms could consider utilizing AI techniques, such as toxic language detection models \cite{toxic}, to flag and make mom vloggers aware of unwanted speech \cite{toxic}.
% , since moms' mental state is vulnerable after giving birth \cite{mom:5}.

\subsection{Limitations and Future Work}
There are several limitations of our work. Firstly, we focus on mom vlogging practices in China and do not claim to generalize our findings to other cultural contexts. Nevertheless, we use mom vlogging as a lens to understand the unique needs and challenges of Chinese moms, and aim to inspire more research in designing information technology for this population. Future work could consider comparing mom vlogging practices in different cultures where women and moms are entitled to different levels of rights, legally and socially. Secondly, given the exploratory nature of this study and the relatively small sample size, the generalizability of the findings is limited. More interviews or a large-scale survey study is needed to obtain a more in-depth understanding of mom vlogging practices. 
% A longitudinal study could also be conducted to understand how mom vloggers' practices evolve over time.
Last but not least, there could be bias in the interview data. Social desirability bias or selective memory is inevitable in the participants' self-reported motivations and experiences of mom vlogging. Such bias could impact the accuracy and reliability of the findings. Though we used the content analysis as a complement to the interviews, future work could conduct a more comprehensive observational study to obtain objective data.

\section{Conclusion}
In this paper, we identify motivations for mom vlogging, i.e., making money, recording life, and seeking identities and values, as well as challenges mom vloggers meet, i.e., video visibility, intensive motherhood and digital work, privacy and self-presentation, personal attacks, and platform regulation. 
We further propose design implications to address the challenges expressed by mom vloggers.
While previous studies on mom blogging have mostly been situated in the Western context, we present an exploratory investigation of mom vloggers in China, which could serve as a lens to understand their unique challenges as women and moms, and encourage more research and design toward promoting their welfare.

\bibliographystyle{plain}
\bibliography{references}  

\appendix

\newpage

\section{Interview Protocol}
\label{interview}
% (This has been lightly edited for length and privacy.)

\noindent \textbf{Introduction}

[REDACTED]

\noindent \textbf{Personal Information}
\begin{itemize}
    \item How old are you? How old is your kid? How many people are in your family?
    \item What is your educational background?
    \item Do you have other jobs other than mom vlogging?
\end{itemize}

\noindent \textbf{Motherhood Life}
\begin{itemize}
    \item When did you leave your paid employment? How did you make the decision? Why did you, instead of your husband, leave the paid employment?
    \item Has your family income reduced after you became a stay-at-home mom? To what extent?
    \item Are there changes in your mentality? Do you feel insecure?
    \item Could you describe a typical day of you?
    \item What are your hobbies?
    \item Do you hang out with friends?
\end{itemize}

\noindent \textbf{Mom Vlogging}
\begin{itemize}
    \item When did you start mom vlogging? What motivated you?
    \item How do you feel about being a mom vlogger? How does your family view your mom vlogging career?
    \item What did you post when you first started mom vlogging? Could you give me an example?
    \item What do you post recently? Do you post good experiences or bad experiences? What was your most recent vlog about? How did the audience react to it?
    \item What challenges do you encounter as a mom vlogger? Could you give me an example?
    \item How do you feel about exposing your personal and family life to the public? How do you feel if your acquaintances see your vlogs?
    \item Do you know other mom vloggers? Do you talk to them?
    \item How do you balance mom vlogging and housework/childrearing?
    \item Do you make money through mom vlogging? How? Does the platform take a part of your revenue? 
\end{itemize}

\end{document}